\documentclass[10pt,twocolumn,twoside]{IEEEtran}
\usepackage{graphicx}
\usepackage{subfigure}
\usepackage{color}
\usepackage{multicol}
\usepackage{amssymb}
\usepackage{multirow}
\usepackage{amsmath}
\usepackage{bm}
\usepackage{cite}
\usepackage{gensymb}
\usepackage{amsfonts}
\usepackage{mathrsfs}
\usepackage{amsmath}
\usepackage{algorithm}
\usepackage{algorithmic}
\usepackage{amsthm}

\usepackage{tabularx}
\usepackage{balance}
\usepackage{mathrsfs}
\usepackage{array}
\usepackage{multirow}
\newcommand{\PreserveBackslash}[1]{\let\temp=\\#1\let\\=\temp}
\newcolumntype{C}[1]{>{\PreserveBackslash\centering}p{#1}}
\newcolumntype{L}[1]{>{\PreserveBackslash\raggedright}p{#1}}
\newcolumntype{R}[1]{>{\PreserveBackslash\raggedleft}p{#1}}
\usepackage[flushleft]{threeparttable}
\begin{document}

	\title{Continuous-Time Channel Prediction Based on Tensor Neural Ordinary Differential Equation}
	\author{\IEEEauthorblockN{ 
Mingyao Cui, Hao Jiang, Yuhao Chen, Yang Du, and Linglong Dai}
	\thanks{
				Mingyao Cui, Hao Jiang, Yuhao Chen, and Linglong Dai are with the Beijing National Research Center for Information Science and Technology (BNRist) as well as the Department of Electronic Engineering, Tsinghua University, Beijing 100084, China (e-mails: \{cmy20, jiang-h18, chen-yh21\}@mails.tsinghua.edu.cn;  daill@tsinghua.edu.cn).
		
				Yang Du is with Huawei Technologies Company Ltd., China. (e-mail: duyang22@huawei.com)
				
		This work was supported in part by the National Key Research and Development Program of China (Grant No. 2020YFB1805005), in part by the National Natural Science Foundation of China (Grant No. 62031019), and in part by the European Commission through the H2020-MSCA-ITN META WIRELESS Research Project under Grant 956256.
}
}
	% \author{
	% ***
% }

	\maketitle
	\IEEEpeerreviewmaketitle
	\begin{abstract}
	Channel prediction is critical to address the channel aging issue in mobile scenarios. Existing channel prediction techniques are mainly designed for discrete channel prediction, which can only predict the future channel in a fixed time slot per frame, while the other intra-frame channels are usually recovered by interpolation. 
However, these approaches suffer from a serious interpolation loss, especially for mobile millimeter-wave communications. To solve this challenging problem, we propose a tensor neural ordinary differential equation (TN-ODE) based continuous-time channel prediction scheme to realize the direct prediction of intra-frame channels. Specifically, inspired by the recently developed continuous mapping model named neural ODE in the field of machine learning, we first utilize the neural ODE model to predict future continuous-time channels. To improve the channel prediction accuracy and reduce computational complexity, we then propose the TN-ODE scheme to learn the structural characteristics of the high-dimensional channel by low-dimensional learnable transform. Simulation results show that the proposed scheme is able to achieve higher intra-frame channel prediction accuracy than existing schemes.
	\end{abstract}
	
	\begin{IEEEkeywords}
		Channel prediction; millimeter-wave communications; massive multiple-input-multiple-output; ordinary differential equation.\vspace{-3mm}
	\end{IEEEkeywords}	

\section{Introduction}
\label{Introduction}
Millimeter-wave (mmWave) massive multiple-input multiple-output (MIMO) has been a critical technology for boosting data transmission speed in 5G communication networks~\cite{01}. By deploying a large number of antennas at the base station (BS), massive MIMO can achieve several orders of magnitude improvements in beamforming gain~\cite{02}. To fully realize this potential, accurate channel state information (CSI) is required at the BS for the efficient design of precoding. According to the current 5G standard~\cite{03}, each frame in 5G wireless communication systems contains multiple time slots, while only the first time slot of each frame is used to estimate the CSI through the predefined sounding reference signal (SRS). Then, the subsequent time slots within the same frame perform precoding design according to the CSI estimated in the first slot.

However, since the channel is time varying in mobile scenarios, the CSI in the first time slot may significantly differ from the actual channels in the subsequent time slots. This is called channel aging in the literature~\cite{05}. Specifically, the channel coherence time is inversely proportional to the carrier frequency and user mobile speed, which could be shorter than the channel estimation period or SRS period in mobile scenarios. For example, for the case of 28 GHz carrier frequency and 60 km/h user mobile speed, the channel coherence time is about 0.32 ms, which is smaller than the shortest SRS period of 0.625 ms defined by the 5G standard~\cite{03}. In this case, the estimated CSI in the first time slot becomes outdated, which could cause a serious spectral efficiency loss of about 30\%~\cite{04}. Therefore, the channel aging problem has to be carefully addressed to enable fast user mobility in mmWave massive MIMO systems.

\subsection{Prior Works}
\label{Prior Works}

To address the channel aging problem, channel prediction techniques have been widely studied to predict the future channels by exploring the channel correlation in the time domain~\cite{04,AR,STAR,FC,RNN,LSTM,RIS-LSTM,Transformer}. There are two typical categories of channel prediction techniques, i.e., model-based and data-based channel prediction. For the first category~\cite{04,AR,STAR}, some classical models are utilized to characterize the time-varying channels, such as the linear extrapolation model~\cite{04}, the auto-regressive (AR) model~\cite{AR}, and the spatio-temporal auto-regressive (ST-AR) model~\cite{STAR}. However, since the actual mobile channels simultaneously suffer from the multi-path effect and the Doppler effect, the time-varying characteristics of actual channels are complicated. Thus, for this category of channel prediction techniques, the fossilized models are difficult to match the time-varying channels, resulting in the unreliable performance in mobile scenarios.

To deal with this problem, data-based channel prediction techniques have been recently proposed to match the time-varying channels in the data-driven way~\cite{FC,RNN,LSTM,RIS-LSTM,Transformer}. Since the neural network models are able to learn the intrinsic complicated feature from data, which could be exploited to improve the channel prediction accuracy. Specifically, in~\cite{FC}, a fully-connected (FC) network was utilized to predict future channels according to the input of high-dimensional historical channels. Then, to decrease the training complexity caused by high-dimensional historical inputs, the recurrent neural network (RNN) like architectures, such as RNN, gate recurrent unit (GRU), and long-short term memory (LSTM), were trained to iteratively process historical channels~\cite{RNN,LSTM,RIS-LSTM}. Furthermore, to avoid the prediction error propagation problem of the sequential prediction of future channels, the transformer model was used to predict future channels in parallel in~\cite{Transformer}.

However, the existing channel prediction techniques~\cite{04,AR,STAR,FC,RNN,LSTM,RIS-LSTM,Transformer} were designed for discrete channel prediction,
 while they fail to directly predict the channels in all time slots of each frame.
To be more specific, as we discussed before, the channels can only be estimated in the first time slot of each frame through the transmission of SRS. Based on these discretely estimated historical channels, the future channels with the same time interval are predicted by existing channel prediction techniques. Then, the channels in other time slots between two adjacent SRS could be recovered by using interpolation methods. Unfortunately, there exists a serious interpolation loss for these discrete channel prediction techniques in mobile scenarios. One possible solution is continuous-time channel prediction for all time slots of each frame. Unfortunately, to the
best of our knowledge, none of the existing methods can achieve continuous-time channel prediction.

\subsection{Our Contributions}
\label{Our Contributions}
To fill in this gap,  we propose a tensor neural ordinary differential equation (TN-ODE) based continuous-time channel prediction scheme in this paper. 
Specifically, inspired by the recently developed continuous-time signal processing technology named neural ODE in the field of machine learning~\cite{ODE}, we adopt the neural ODE architecture proposed in~\cite{ODE} to model the continuous-time channel prediction problem. In the above architecture, a GRU-based encoder is used to preprocess the discretely sampled historical channels, then a neural ODE-based decoder is used to predict future channels in consecutive time slots. Furthermore, to improve the channel prediction accuracy and reduce the computational complexity of the neural ODE, we propose the TN-ODE to exploit the structural characteristics of channels in multiple domains by a series of low-dimensional learnable transforms. To be more specific, in the antenna domain, the channel model is described by different angles of arrival (AoAs) and angles of departure (AoDs), while in the frequency domain, the channel model is mainly determined by multiple times of arrival (ToAs). Thanks to these structural characteristics, the proposed TN-ODE allows us to decouple the complicated high-dimensional channel prediction into efficient low-dimensional channel prediction in multiple domains. Simulation results show that the proposed TN-ODE based continuous-time channel prediction technique can effectively mitigate the interpolation loss and improve the channel prediction performance in all time slots of each frame.

\subsection{Organization and notation}
\label{Organization and notation}
The remainder of this paper is organized as follows. In Section \ref{System model}, the system model of the mmWave massive MIMO is introduced, and the continuous-time channel prediction problem in this system is then formulated. After that, we elaborate on the proposed TN-ODE based continuous-time channel prediction model in Section \ref{Proposed Method}. Section \ref{SIMULATION-RESULTS} illustrates the simulation results. Finally, conclusions are drawn in Section \ref{CONCLUSION}.

{\it Notation}: We denote the column vector $\bf a$ and matrix $\bf A$ by boldface lower-case and upper-case letters, respectively; ${\bf A}^T$, ${\bf A}^H$, and ${\bf A}^{-1}$ are the transpose, conjugate transpose, and inverse of the matrix $\bf A$, respectively; ${\bf A}\otimes {\bf B}$ is the Kronecker product of the matrix $\bf A$ and matrix $\bf B$; {\color{black} ${\bf A}\circ {\bf B}$ is the Hadamard product of $\bf A$ and $\bf B$};  ${\bf I}_{N}$ denotes an $N\times N$ identity matrix. $\mathcal{CN}\left({\mu,\sigma^2}\right)$ is the probability density function of the circularly symmetric complex Gaussian distribution with mean $\mu$ and variance $\sigma^2$. $\mathbb{E}\left\lbrace\cdot\right\rbrace$ denotes the statistical expectation. { We use $\text{vec}(\mathbf{A})$ to denote the vectorization of matrix $\mathbf{A}$. $\sigma(x) = \frac{1}{1 + e^{-x}}$ and $\tanh(x) = \frac{e^x - e^{-x}}{e^x + e^{-x}}$ represent the Sigmoid function and hyperbolic tangent function, respectively. We denote $h[n], n\in\mathbb{Z}$ as a discrete-time sequence and $h(t), t\in\mathbb{R}$ as a continuous-time sequence.}

\section{System model}
\label{System model}

In this section, we will first introduce the signal model of the mmWave massive MIMO system. Then, the continuous-time channel prediction is formulated to avoid the interpolation loss problem in existing discrete channel prediction schemes.

	\begin{figure}
	\centering 
	%		\vspace*{-0.5em}
	\includegraphics[width=3.3in]{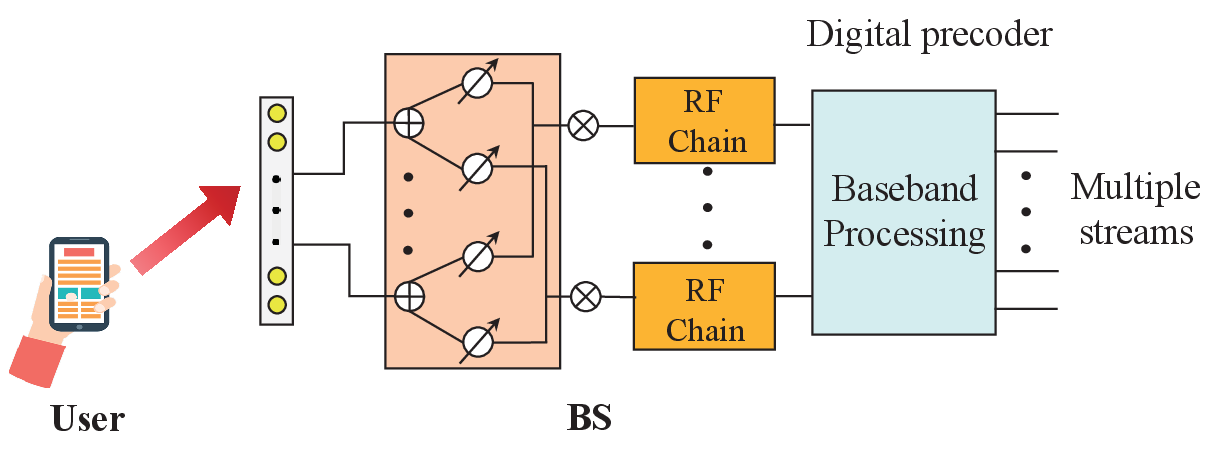}
	\caption{  Hybrid precoding for mmWave massive MIMO.
	} 
	\label{fig1}
	\vspace*{-1em}
\end{figure}	

% \begin{figure}[tp]
% 	\begin{center}
% 		\hspace*{0mm}\includegraphics[width=1\linewidth]{system.eps}
% 	\end{center}
% 	\vspace*{-4mm}
% 	\caption{Hybrid precoding for mmWave massive MIMO.} 
% 	\label{fig1}
% 	\vspace*{+1mm}
% \end{figure} 

% \begin{figure}[tp]
% 	\begin{center}
% 		\hspace*{0mm}\includegraphics[width=1\linewidth]{frame.eps}
% 	\end{center}
% 	\vspace*{-4mm}
% 	\caption{The 5G frame structure defined by 3GPP~\cite{03}.} 
% 	\label{fig2}
% 	\vspace*{+1mm}
% \end{figure} 

	\begin{figure}
	\centering 
	%		\vspace*{-0.5em}
	\includegraphics[width=3.3in]{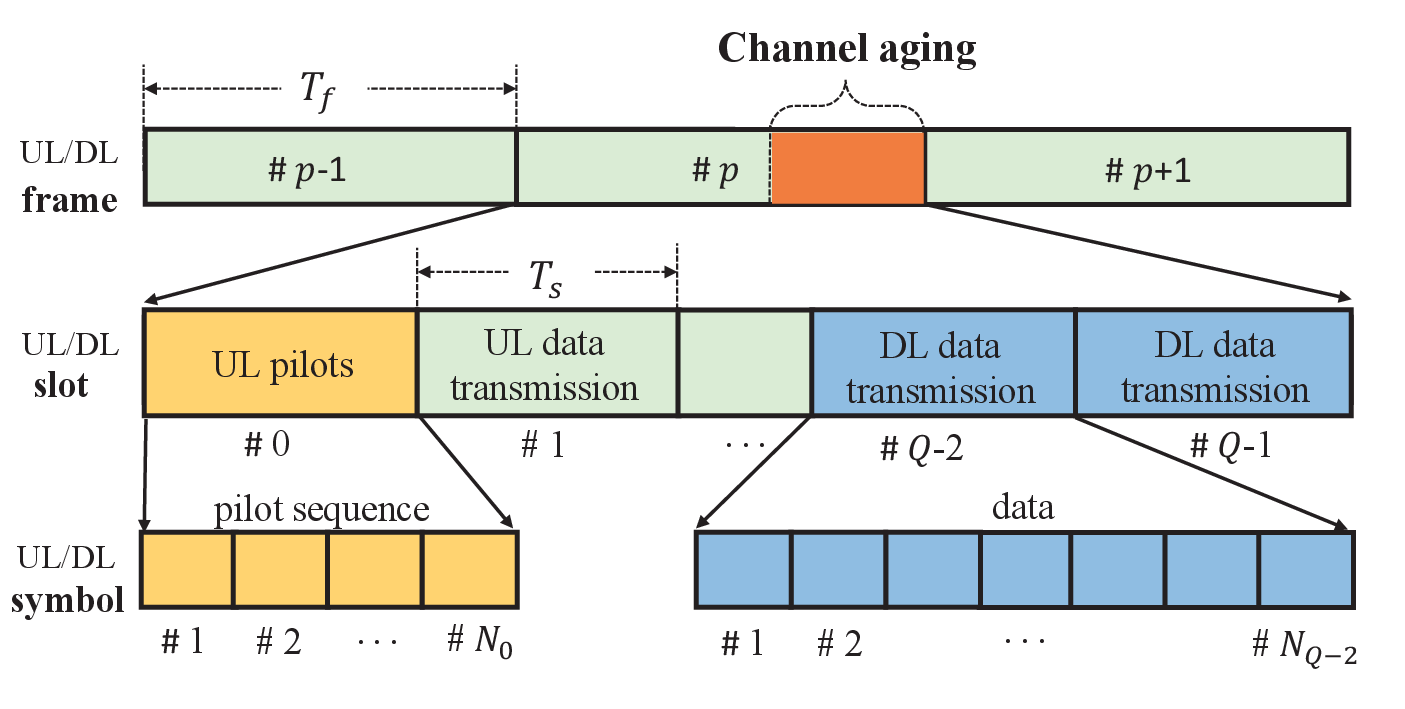}
	\caption{  The 5G frame structure defined by 3GPP~\cite{03}.
	} 
	\label{fig2}
	\vspace*{-1em}
\end{figure}

\subsection{Signal model}
\label{Signal model}

In this paper, we consider an uplink time division duplexing (TDD) based mmWave massive MIMO system with orthogonal frequency division multiplexing (OFDM). The base station (BS) equipped with an $N_{\rm T}$-antenna uniform linear array (ULA)~\cite{ULA} serves an $N_{\rm R}$-antenna user with $M$ subcarriers. To reduce energy consumption, hybrid precoding structure is employed in the BS~\cite{hybrid}, where the number of radio frequency (RF) chains is $N_{\rm RF}$, as indicated in Figure \ref{fig1}.
% where $N_{\rm RF}N_{\rm T}$ phase shifts with low energy consumption are used to decrease . 
According to the 5G standard~\cite{03}, the time resources for communication are divided into frames and each frame consists of $Q$ time slots. As shown in Figure \ref{fig2}, the $Q$ time slots could be further divided into three parts, i.e., uplink pilots, uplink data transmission, and downlink data transmission. For the $q$-th time slot, $N_q$ symbols are included and the channel remains unchanged during the $N_q$ symbols based on the block-fading assumption~\cite{blockfading}, where the channel remains time-invariant within each time slot and changes between different time slots.

Let ${{\bf H}_m{(t)}\in\mathbb{C}^{N_{\rm T}\times N_{\rm R}}}$ denote the channel at the time $t$. Due to the limited number of scattering clusters in the mmWave propagation environment, we adopt the widely used geometric Saleh-Valenzuela multipath channel model~\cite{hybrid} to characterize the mmWave channel. Under this model, ${\bf H}_m{(t)}$ can be denoted as
\begin{align}\label{eq3}
{\bf H}_m{(t)}=\sum_{l=1}^L\alpha_l e^{-j2\pi (v_l t + f_m \tau_l)}{\bf a}_{\rm T}(\phi_{l,{\rm T}}){\bf a}_{\rm R}^H(\phi_{l,{\rm R}}),
\end{align}
where $L$ is the number of the paths, $\alpha_l$, $v_l$, $\tau_l$, $\phi_{l,{\rm T}}$, and $\phi_{l,{\rm R}}$ are the complex path gain, Doppler shift, ToA, AoA, and AoD of the $l$-th path, respectively. For $m \in \{1,2,\cdots, M\}$, $f_m = f + \frac{B}{2}(m - \frac{M}{2})$ denotes the $m$-th subcarrier frequency, with $f$, $B$, and $M$ being the carrier frequency, bandwidth, and the number of subcarriers.
Since the ULA is considered in this paper, the array steering vector ${\bf a}_{\rm T}(\phi_{l,{\rm T}})$ and ${\bf a}_{\rm R}(\phi_{l,{\rm R}})$ could be represented by
\begin{align}\label{eq4}
{\bf a}_{\rm T}(\phi_{l,{\rm T}})&=\frac{1}{\sqrt{N}}[e^{-j\frac{2\pi}{\lambda}d{\rm sin}(\phi_{l,{\rm T}}){\bf n}_{\rm T}}],\\
{\bf a}_{\rm R}(\phi_{l,{\rm R}})&=\frac{1}{\sqrt{M}}[e^{-j\frac{2\pi}{\lambda}d{\rm sin}(\phi_{l,{\rm R}}){\bf n}_{\rm R}}],
\end{align}
where ${\bf n}_{\rm T}=[0,1,\cdots,N_{\rm T}-1]^T$ and ${\bf n}_{\rm R}={[0,1,\cdots,N_{\rm R}-1]^T}$,  $\lambda$ is the carrier wavelength, and $d$ is the antenna spacing usually set as $d=\lambda/2$.

We denote $T_f$ and $T_s$ as the duration time of one frame and one time slot, where $T_f = Q T_s$. Accordingly, we can use ${\bf H}_m^{(p,q)} = {\bf H}_m{(t_{p,q})}$ to denote the channel at the $q$-th time slot of the $p$-th frame and the $m$-th subcarrier, where $t_{p,q} = {pT_f+qT_s}$.
Then, the received signal ${{\bf Y}_m^{(p,q)}\in\mathbb{C}^{N_{\rm RF}\times N_q}}$ at the $q$-th time slot of the $p$-th frame and the $m$-th subcarrier in BS can be expressed by
\begin{align}
{\bf Y}_m^{(p,q)}&={\bf A}^{(p,q)}{\bf H}_m^{(p,q)}{\bf S}_m^{(p,q)}+{\bf A}^{(p,q)}{\bf N}_m^{(p,q)}\notag\\
&={\bf\overline H}_m^{(p,q)}{\bf S}_m^{(p,q)}+{\bf A}^{(p,q)}{\bf N}_m^{(p,q)},\label{eq12}
\end{align}
where ${{\bf A}^{(p,q)}\in\mathbb{C}^{N_{\rm RF}\times N_{\rm T}}}$ is the frequency-independent combining matrix, ${{\bf S}_m^{(p,q)}\in\mathbb{C}^{N_{\rm R}\times N_q}}$ denotes the transmitted signal, ${{\bf N}_m^{(p,q)}\in\mathbb{C}^{N_{\rm RF}\times N_q}}$ is the Gaussian noise and each element following the distribution $\mathcal{CN}(0,\sigma^2)$ with $\sigma^2$ being the noise power, and ${{\bf\overline H}_m^{(p,q)}\in\mathbb{C}^{N_{\rm RF}\times N_{\rm R}}}$ is the effective channel matrix in the $q$-th time slot of the $p$-th frame and the $m$-th subcarrier.

We utilize the discrete Fourier transmission (DFT) codebook to design the analog combining ${\bf A}^{(p,q)}$~\cite{DFT}. In the DFT codebook, each codeword points to a specific azimuth AoA and all codewords will cover the entire beam space. By traversing all codewords, the strongest ${N_{\rm RF}}$ codewords could be selected to construct ${\bf A}^{(p,q)}$. Benefiting from the fact that the time-varying channel is mainly caused by the Doppler effect, while the AoA and AoD are time-invariant in several frames during tens of milliseconds~\cite{beaminvariant}, the optimal combining matrix stays unchanged in several frames. In this case, we suppose  ${\bf A}^{(p,q)}={\bf A},\forall p\in\{0,1,\cdot\cdot\cdot,P-1\},\forall q\in\{0,1,\cdot\cdot\cdot,Q-1\}$, where $P$ is the number of frames in the order of tens of milliseconds.

{\color{black}In particular, when $q=0$, the effective channel ${\bf\overline H}_m^{(p,0)}$ of the first time slot in the $p$-th frame is estimated according to the predefined pilot sequence ${\bf S}_m^{(p,0)}$ and received signal  ${\bf Y}_m^{(p,0)}$. Generally, we use the least square (LS) channel estimation method to recover the effective channel, which could be represented by
\begin{align}\label{eq2}
{\rm vec}({\bf\hat H}_m^{(p,0)})=({\bf S}_m^{{(p,0)}^T}\otimes{\bf I}_{{N}_{\rm RF}})^{-1}{\rm vec}({\bf Y}_m^{(p,0)}),
\end{align}
where ${\rm vec}({\bf\hat H}_m^{(p,0)})$ is the vectorization of LS channel estimation ${\bf\hat H}_m^{(p,0)}$. }When $q\neq0$, the ${{\bf S}_m^{(p,q)}}$ is the transmitted signal and the achievable average rate $R$ could be written as 
\begin{align}\label{eq31}
&R= \\
&\frac{1}{M}\sum_{m=1}^M{\rm log}_2\left|{\bf I}_{N_{\rm R}}+ \frac{1}{N_{\rm R}\sigma^2}{\bf D}_m^{(p,q)}{\bf\overline H}_m^{(p,q)}{\bf\overline H}_{m}^{{(p,q)}^H}{\bf D}_{m}^{{(p,q)}^H} \right|. \notag
\end{align}
We utilize the classical zero-forcing method~\cite{02} to design the digital precoding ${\bf D}_m^{(p,q)}\in\mathbb{C}^{N_{\rm R}\times N_{\rm RF}}$ in the $q$-th time slot of the $p$-th frame and the $m$-th subcarrier as:
\begin{align}\label{eq32}
{\bf D}_m^{(p,q)}=({\bf\hat H}_{m}^{{(p,q)}^H}{\bf\hat H}_m^{(p,q)})^{-1}{\bf\hat H}_{m}^{{(p,q)}^H}.
\end{align}

\subsection{Problem formulation}
\label{Problem formulation}
To calculate the digital precoding ${\bf D}_m^{(p,q)}$, the estimated instantaneous ${\bf\hat H}_m^{(p,q)}$ is required according to (\ref{eq32}). Whereas, since only the CSI at the first time slot of each frame ${\bf\hat H}_m^{(p,0)}$ is available, we usually use the ${\bf\hat H}_m^{(p,0)}$ to perform precoding for the subsequent time slots, i.e., ${\bf D}_m^{(p,q)}={\bf D}_m^{(p,0)}$. Unfortunately, due to the channel aging issue induced by mobility, the outdated CSI ${\bf\hat H}_m^{(p,0)}$ has a significant change compared with the actual effective channel ${\bf\overline H}_m^{(p,q)}$, which results in a sever performance loss for mmWave MIMO in mobile scenarios. 

To mitigate the performance loss caused by channel aging, some channel prediction techniques~\cite{04,AR,STAR,FC,RNN,LSTM,RIS-LSTM,Transformer} have been proposed to deal with the channel aging issue by exploring the temporal correlation of the time-varying channel. Specifically, the existing channel prediction schemes could predict future channels in discrete frames, i.e., ${\bf\hat H}_{m}^{{(p+1,0)}},\cdot\cdot\cdot,{\bf\hat H}_{m}^{{(p+K,0)}}$, based on the historical channels ${\bf\hat H}_{m}^{{(p-J,0)}},\cdot\cdot\cdot,{\bf\hat H}_{m}^{{(p,0)}}$ with the same time interval. Since these channel prediction methods are designed for discrete channel prediction, which only predict the channel in the first time slot of each frame, they can not realize the direct prediction of the channels for all time slots in future frames. Thus, the interpolation method has to be utilized to recover the channels ${\bf\hat H}_{m}^{{(p+k,q)}}$ with $q>0$ as
\begin{align}\label{eq5}
{\bf\hat H}_{m}^{{(p+k,q)}}=(1-\frac{q}{Q}){\bf\hat H}_{m}^{{(p+k,0)}}+\frac{q}{Q}{\bf\hat H}_{m}^{{(p+k+1,0)}},
\end{align}
where $k=0,1,\cdot\cdot\cdot,K-1$. However, due to the complicated change of the channel, simple interpolation is difficult to describe the actual change of the channel. Therefore, there is an interpolation loss for the existing discrete channel prediction schemes.

Unlike the existing discrete channel prediction schemes, we reformulate the channel prediction problem as a continuous-time channel mapping problem to avoid interpolation loss. Specifically, we utilize the historical discrete channels from the past $J$ frames to predict the future continuous-time channels in the next $K$ frames, which could be formulated as 
\begin{subequations}
\begin{align}
		&\mathop{\rm min}\limits_{\bm{\theta}}\ \sum\limits_{k=0}\limits^{K-1}\sum\limits_{q=0}\limits^{Q-1}\sum\limits_{m=1}\limits^{M}\mathbb{E}\bigg\{\frac{\|{\bf\overline H}_m^{(p+k,q)}-{\bf\hat H}_m^{(p+k,q)}\|^2}{\|{\bf\overline H}_m^{(p+k,q)}\|^2}\bigg\},\label{eq6:sub1}  \\ 
	 &\ {\rm s. t.} \ \ ({\bf\hat H}_m^{(p,1)},{\bf\hat H}_m^{(p,2)},\cdot\cdot\cdot,{\bf\hat H}_m^{(p+K-1,Q-1)})\notag\\
	 &\ \ \ \ \ \ \ \ \ \ \ \ \ \ \ \ \ \ \ \ \ \ \ \ \ =f({\bf\hat H}_m^{(p-J,0)},\cdot\cdot\cdot,{\bf\hat H}_m^{(p,0)}; {\bm \theta}),\label{eq6:sub2}
\end{align}\label{eq6}
\end{subequations}
\!\!where $f(\cdot)$ is the proposed continuous-time channel prediction model and ${\bm \theta}$ is the parameters of the model. Since the normalized mean square error (NMSE) is not affected by the amplitude of the channel, we adopt the NMSE as the minimization target to realize stable convergence. It is worth noting that the estimated historical channels are discretely sampled at the first time slot of each frame. Correspondingly, the predicted channels are continuously distributed at any time slot of each future frame. By contrast, the existing discrete channel prediction schemes only predict the channel at the first time slot of the future frames. Thus, the proposed continuous-time channel prediction scheme realizes the direct prediction of the future channel in any time slot so that the interpolation loss can be avoided.

{\color{black}
\section{Proposed Method}
\label{Proposed Method}
In this section, we first introduce the background of neural ODE and elaborate on the framework of neural ODE based channel prediction. Then, we propose the TN-ODE to explore the  mmWave channel structure to improve the channel prediction performance.

	\begin{figure*}
	\centering 
	%		\vspace*{-0.5em}
	\includegraphics[width=7in]{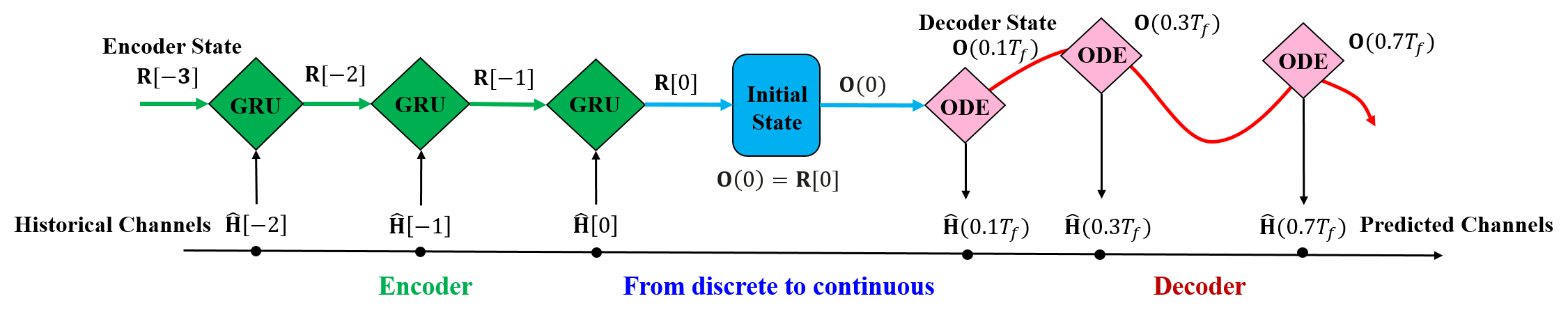}
	\caption{  The framework of neural ODE-based channel prediction.
	} 
	\label{fig:framework}
	\vspace*{-1em}
\end{figure*}

% \begin{figure*}[tp]
% 	\begin{center}
% 		\hspace*{0mm}\includegraphics[width=1\linewidth]{ODE.png}
% 	\end{center}
% 	\vspace*{-4mm}
% 	\caption{The framework of neural ODE-based channel prediction.} 
% 	\label{fig:framework}
% 	\vspace*{+1mm}
% \end{figure*} 

\subsection{Background of Neural ODE}
To achieve continuous-time channel prediction, it is crucial to find an appropriate technique to process continuous-time signals. Recently, with the rapid advancement in the field of dynamical systems, neural ODE becomes an attractive technology for modeling continuous-time sequences~\cite{ODE, NODE, LatentODE}. Neural ODEs use first-order differential equations to fit the hidden state of time sequences, so it is capable of handling continuous-time signals. 
To make this paper self-contained, we provide a brief background of neural ODE. Specifically, classical RNN-like architectures, including RNN, GRU, and LSTM, build complicated networks to encode time sequences into a series of hidden states:
\begin{align}\label{eq:DDE}
 \mathbf{h}[n] = \mathbf{h}[n-1] + g(\mathbf{h}[n-1], \bm{\theta}).
\end{align}
Here, $\mathbf{h}[n]$ represents the hidden state at the $n$-th discrete time, $g(\cdot)$ denotes the state transition function realized by neural networks, and $\bm{\theta}$ is the network parameters. The transition in (\ref{eq:DDE}) is built on a discrete difference equation, which is awkward to deal with signals not belonging to discrete time samples.  On the contrary, neural ODEs define a continuous-time hidden state $\mathbf{h}(t)$, which can be formulated as a time-invariant differential equation: 
\begin{align}\label{eq:ODE1}
 \frac{\text{d}\mathbf{h}(t)}{\text{d} t} = f(\mathbf{h}(t), \bm{\theta}).
\end{align}
Besides, (\ref{eq:ODE1}) is equivalent to the following integral form: 
\begin{align}\label{eq:ODE2}
 \mathbf{h}(t) = \int_{t_0}^t f(\mathbf{h}(\tau), \bm{\theta}) \text{d} \tau + \mathbf{h}(t_0).
\end{align}
Here, $\mathbf{h}(t_0)$ is the initial hidden state, and function $f(\mathbf{h}(t), \bm{\theta})$ describes the dynamic of hidden state $\mathbf{h}(t)$.
One can acquire the hidden state $\mathbf{h}(t)$ at an arbitrary time $t$ by solving problem (\ref{eq:ODE2}) through an ODE solver:
\begin{align} \label{eq:ODESolver}
	\mathbf{h}(t) = \text{ODESolver}(f(\cdot, \bm{\theta}), \mathbf{h}(t_0), t_0, t).
\end{align}
As indicated in \cite{NODE}, such an ODE solver can be implemented by various numerical schemes, including the forward and backward Euler methods, the Runge-Kutta method, and the linear multi-step method. 
As a consequence, applying neural ODE models (\ref{eq:ODE2}) and solvers (\ref{eq:ODESolver}) allows us to deal with continuous-time sequences, so as to achieve continuous-time channel prediction. 
% Based on this model, one can use 

\subsection{Framework of Neural ODE Based Channel Prediction}

Based on the above background, the framework of neural ODE-based channel prediction is presented in this subsection. Our aim is to predict the channels for all time slots of the future $K$ frames by processing those historical $J$ channels. The Latent ODE architecture introduced in \cite{LatentODE} is adopted to model this process. 
For expression clarity, we denote 
$\hat{\mathbf{H}}[n] = [\text{vec}(\hat{\mathbf{H}}_1^{(n,0)}), \text{vec}(\hat{\mathbf{H}}_2^{(n,0)}),  \cdots, \text{vec}(\hat{\mathbf{H}}_M^{(n,0)}) ] $
and 
$\hat{\mathbf{H}}(t) = [\text{vec}(\hat{\mathbf{H}}_m{(t)}), \text{vec}(\hat{\mathbf{H}}_2{(t)}),  \cdots, \text{vec}(\hat{\mathbf{H}}_M{(t)}) ]$.

 As shown in Figure \ref{fig:framework}, the neural ODE-based channel prediction is composed of two modules, i.e., an encoder and a decoder~\cite{LatentODE}. Generally speaking, the encoder is responsible for extracting features from the historical channels $\hat{\mathbf{H}}[n]$ for $n = \{0, -1, \cdots, -J + 1\}$. The output of the encoder serves as the initial state of the decoder. Correspondingly, the decoder exploits a neural ODE to infer future continuous-time channels $\hat{\mathbf{H}}(t)$ for $t >0$.  

Specifically, the encoder's role is to extract the features from historical channels. Since the SRS signals are transmitted and received with equally-sized time interval $T_f$,  RNN-like architectures are enough to deal with these sequences. We denote the hidden state of $\hat{\mathbf{H}}[n]$ as $\mathbf{R}[n]$. Then, based on the Markov property of RNN models, the map from $\mathbf{R}[n-1]$ to  $\mathbf{R}[n]$ can be written as 
\begin{align}\label{eq:Encoder}
	\mathbf{R}[n] = \text{EncoderCell}(\mathbf{R}[n-1], \hat{\mathbf{H}}[n], \bm{\theta}_E),
\end{align}
where $\text{EncoderCell}(\cdot)$ is the transition function of the RNN-like network with $\bm{\theta}_E$ being the learnable parameters. 

For the decoder, a neural ODE model is deployed to specific the dynamic of the future channel's hidden state. This hidden state is defined as $\mathbf{O}(t)$. Besides, the final output $\mathbf{R}[0]$ of the encoder is regarded as the initial state $\mathbf{O}(0)$ of decoder. Therefore, for any times $t > 0$, the hidden state $\mathbf{O}(t)$ can be presented as 
\begin{align}\label{eq:Decoder}
	\frac{\text{d}\mathbf{O}(t)}{\text{d}t} = \text{DecoderCell}(\mathbf{O}(t), \bm{\theta}_D) ,
\end{align}
where $\text{DecoderCell}(\cdot)$ denotes the transition function of the neural ODE network with $\bm{\theta}_D$ being its learnable parameters. Note that $(\ref{eq:Decoder})$ can be solved by the ODESolver in (\ref{eq:ODESolver}). After that, one layer neural network $\text{Pred}(\cdot)$ is built to output the predicted channel $\hat{\mathbf{H}}(t)$ from the hidden state $\mathbf{O}(t)$:
\begin{align}
	\hat{\mathbf{H}}(t) = \text{Pred}(\mathbf{O}(t), \bm{\theta}_P),
\end{align}
where $\bm{\theta}_P$ denotes its parameters. All in all, following this neural ODE framework, we are capable of extracting features from the previous channels and then predicting future continuous-time channels for any $t$. 

% ical RNN-like architectures, including RNN, GRU, and LSTM, usually use discrete different equations to model time sequences. The   
\subsection{TN-ODE based Channel Prediction}

In this subsection, we elaborate on the idea of tensor neural ODE for designing the three crucial transition functions: $\text{EncoderCell}(\cdot)$, $\text{DecoderCell}(\cdot)$, and $\text{Pred}(\cdot)$. 

We commence our discussion by briefly introducing the transition functions widely used in classical neural ODE framework \cite{LatentODE}. The authors in \cite{LatentODE} deployed a GRU model as its encoder transition function and modified the GRU model to act as the decoder transition function. To fit in our channel prediction framework, the inputs, hidden states, and outputs should be first vectorized as the following column vectors: $\hat{\mathbf{h}}[n] = \text{vec}(\hat{\mathbf{H}}[n])$, $\mathbf{r}[n] = \text{vec}(\mathbf{R}[n])$, $\mathbf{o}(t) = \text{vec}(\mathbf{O}(t))$, and $\hat{\mathbf{h}}(t) = \text{vec}(\hat{\mathbf{H}}(t))$. Then, according to the GRU architecture~\cite{LatentODE}, $\text{EncoderCell}(\cdot)$ consists of the following modules:
\begin{subequations}
\begin{align}
	&\mathbf{z}  = \sigma\left(\mathbf{U}^{z}\hat{\mathbf{h}}[n] + \mathbf{W}^{z} \mathbf{r}[n - 1] \right), \label{eq:GRU1a}\\
	&\mathbf{x}  = \sigma\left(\mathbf{U}^{x}\hat{\mathbf{h}}[n] + \mathbf{W}^{x} \mathbf{r}[n - 1] \right), \label{eq:GRU1b}\\
	&\mathbf{u}  = \tanh \left(\mathbf{U}^{u}\hat{\mathbf{h}}[n] + \mathbf{W}^{u} (\mathbf{r}[n - 1] \circ \mathbf{x}) \right), \label{eq:GRU1c}\\
	&\mathbf{r}[n]  = (\mathbf{1} - \mathbf{z})\circ \mathbf{u} + \mathbf{z} \circ \mathbf{r}[n - 1],\label{eq:GRU1d}
\end{align}
\end{subequations}
where matrices $\{\mathbf{U}, \mathbf{W}\}$ are the learnable parameters. As for the decoder, it is different from the encoder which can receive 
external stimulus $\hat{\mathbf{h}}[n]$ to update its states. The hidden state transition of $\text{DecoderCell}(\cdot)$ is an auto-regressive process without external stimulus. Thereby, to fit GRU model in this decoder, we can carry out the steps below to modify GRU: remove $\hat{\mathbf{h}}[n]$ from (\ref{eq:GRU1a})-(\ref{eq:GRU1d}); replace $\mathbf{r}[n - 1]$ and $\mathbf{r}[n]$ with $\mathbf{o}(t)$ and $\frac{\text{d}\mathbf{o}(t)}{\text{d}t}$, respectively. 
Finally, function $\text{Pred}(\cdot)$ can be realized by a fully connected layer, i.e. $\hat{\mathbf{h}}(t) = \mathbf{W}^h \mathbf{o}(t)$.  
As a result, the entire neural ODE-based channel prediction is successfully established based on the classical GRU model. 

There is no denying that the above transition functions have the ability to process continuous-time sequences. However, they will suffer from two serious problems when applied to channel prediction. First, these transition functions fail to exploit the underlying channel structure. As shown in (\ref{eq3}), mmWave channels exhibit obvious correlations in multiple domains. For example, the antenna-domain channel is constructed by the superposition of multiple array steering vectors with different AoAs and AoDs. Besides, in the frequency domain, the channel structure can be captured by several ToAs. However, simply vectorizing channels to fit in the GRU model will undermine such regular structures. Second, the computational complexity of these transition functions is also unaffordable. Take the function $\text{Pred}(\cdot)$ as an example, we suppose $\mathbf{W}^h$ is a square matrix. As the dimension of $\hat{\mathbf{h}}(t)$ is $N_{\text{RF}} N_{\rm R} M \times 1$, then matrix $\mathbf{W}^h$ will contain $2N_{\text{RF}}^2 N_{\rm R}^2 M^2$ floating points. If $N_{\text{RF}} = N_{\rm R} = 4$ and $M = 256$, then the number of floating points of $\mathbf{W}^h$ is $2N_{\text{RF}}^2 N_{\rm R}^2 M^2 = 33,554,432$, which costs unacceptable computational resources.

To address these two critical problems, we propose the TN-ODE by exploiting the channel correlation. Our scheme is inspired by the tensor decomposition based signal processing algorithms~\cite{TD_Zhou2017}, which extract the information of channels from different domains and process them separately.
In our model, we preserve the matrix form of $\hat{\mathbf{H}}[n]$, $\mathbf{R}[n]$, $\mathbf{O}(t)$, and $\hat{\mathbf{H}}(t)$, and use different learnable transforms to independently extract the antenna-domain and frequency-domain information from historical channels. 
We take the matrix product $\mathbf{U}^{z}\hat{\mathbf{h}}[n]$ in (\ref{eq:GRU1a}) as an example. The classical GRU model (\ref{eq:GRU1a}) vectorizes $\hat{\mathbf{H}}[n]$ as $\hat{\mathbf{h}}[n] \in \mathbb{C}^{N_{\text{RF}}N_{\rm R} M \times 1}$ and uses a high-dimensional matrix $\mathbf{U}^{z}$ to process $\hat{\mathbf{h}}[n]$. Instead, we keep the shape of $\hat{\mathbf{H}}[n]$ as $N_{\text{RF}}N_{\rm R} \times M$ and use two independent low-dimensional matrices $\mathbf{U}^{z}_l$ and $\mathbf{U}^{z}_r$ to separately work on the antenna domain and frequency domain of $\hat{\mathbf{H}}[n]$, which gives rise to $\mathbf{U}^{z}_l \hat{\mathbf{H}}[n] \mathbf{U}^{z}_r$. Similarly, we modify all modules in (\ref{eq:GRU1a})-(\ref{eq:GRU1d}) by the same means to construct the tensor-inspired $\text{EncoderCell}(\cdot)$ as 
\begin{subequations}
\begin{align}
	&\mathbf{Z}  = \sigma\left(\mathbf{U}^{z}_l\hat{\mathbf{H}}[n]\mathbf{U}^{z}_r + \mathbf{W}^{z}_l \mathbf{R}[n - 1]\mathbf{W}^{z}_r \right), \label{eq:GRU2a}\\
	&\mathbf{X}  = \sigma\left(\mathbf{U}^{x}_l\hat{\mathbf{H}}[n]\mathbf{U}^{x}_r + \mathbf{W}^{x}_l \mathbf{R}[n - 1]\mathbf{W}^{x}_r \right), \label{eq:GRU2b}\\
	&\mathbf{U}  = \tanh \left(\mathbf{U}^{u}_l\hat{\mathbf{H}}[n]\mathbf{U}^{u}_r + \mathbf{W}^{u}_l (\mathbf{R}[n - 1] \circ \mathbf{X})\mathbf{W}^{u}_r \right), \label{eq:GRU2c}\\
	&\mathbf{R}[n]  = (\mathbf{1} - \mathbf{Z})\circ \mathbf{U} + \mathbf{Z} \circ \mathbf{R}[n - 1],\label{eq:GRU2d}
\end{align}
\end{subequations}
where matrices $\{\mathbf{U}_l, \mathbf{U}_r, \mathbf{W}_l, \mathbf{W}_r\}$ are all learnable parameters. 
Here, matrices $\{\mathbf{U}_l^z, \mathbf{U}_l^x, \mathbf{U}_l^u\}$ have a size of $F_l \times N_{\text{RF}}N_{\rm R}$, matrices $\{\mathbf{U}_r^z, \mathbf{U}_r^x, \mathbf{U}_r^u\}$ have a size of $M \times F_r$,  matrices $\{\mathbf{W}_l^z, \mathbf{W}_l^x, \mathbf{W}_l^u\}$ have a size of $F_l \times F_l$,  and matrices $\{\mathbf{W}_r^z, \mathbf{W}_r^x, \mathbf{W}_r^u\}$ have a size of $F_r \times F_r$.  $F_l$ and $F_r$ denote the feature dimensions.  {\color{black} Notice that the computations in (\ref{eq:GRU2a})-(\ref{eq:GRU2d}) are all complex-valued multiplications, which are realized by the complex-valued neural network (CVNN) proposed in \cite{STAR}. The Sigmoid and Tanh functions work on the real and imaginary parts respectively.} 
Moreover, we can use the same way to transform the classical $\text{DecoderCell}(\cdot)$ and $\text{Pred}(\cdot)$ to their tensor forms. To be specific, the transition function of the ODE decoder can be written as 
\begin{subequations}
\begin{align}
	&\mathbf{Z}(t)  = \sigma\left(\mathbf{V}^{z}_l \mathbf{O}(t)\mathbf{V}^{z}_r \right), \label{eq:ODEa}\\
	&\mathbf{X}(t)  = \sigma\left(\mathbf{V}^{x}_l \mathbf{O}(t)\mathbf{V}^{x}_r \right), \label{eq:ODEb}\\
	&\mathbf{U}(t)  = \tanh \left(\mathbf{V}^{u}_l (\mathbf{O}(t) \circ \mathbf{X}(t))\mathbf{V}^{u}_r \right), \label{eq:ODEc}\\
	&\frac{\text{d}\mathbf{O}(t)}{\text{d}t}  = (\mathbf{1} - \mathbf{Z}(t))\circ \mathbf{U}(t) + \mathbf{Z}(t) \circ \mathbf{O}(t), \label{eq:ODEd}
\end{align}
\end{subequations}
where the size of matrices $\{\mathbf{V}^{z}_l, \mathbf{V}^{x}_l,\mathbf{V}^{u}_l \}$ is $F_l \times F_l$ and the size of matrices $\{\mathbf{V}^{z}_r, \mathbf{V}^{x}_r,\mathbf{V}^{u}_r \}$ is $F_r \times F_r$. Then, the function of $\text{Pred}(\cdot)$ can be given by 
\begin{align}\label{eq:Pred}
	\hat{\mathbf{H}}(t) = \mathbf{W}^h_l \mathbf{O}(t)  \mathbf{W}^h_r,
\end{align}
with $\mathbf{W}^h_l \in \mathbb{C}^{N_{\text{RF}} N_{\rm R} \times F_l }$ and $\mathbf{W}^h_r \in \mathbb{C}^{F_r \times M }$.

Our proposed TN-ODE enjoys two crucial merits compared to the classical one \cite{LatentODE}. To begin with, it preserves the structural features of multi-domain channels in the entire procedure, so our scheme is specific for predicting wireless continuous-time channels. Moreover, its computational complexity is much lower than that of \cite{LatentODE}. We still take the function $\text{Pred}(\cdot)$ as an example. As shown in (\ref{eq:Pred}), we suppose both $\mathbf{W}^h_l$ and $\mathbf{W}^h_r$ are square matrices. Since the shape of $\hat{\mathbf{H}}(t)$ is $N_{\text{RF}}N_{\rm R} \times M$, matrices $\mathbf{W}^h_l$ and $\mathbf{W}^h_r$ have sizes of $N_{\text{RF}}N_{\rm R} \times N_{\text{RF}}N_{\rm R}$ and $M \times M$, which gives rise to $2(N_{\text{RF}}^2N_{\rm R}^2 + M^2)$ floating points. Therefore, if $N_{\text{RF}} = N_{\rm R} = 4$ and $M = 256$, the number of floating points is decreasing from $33,554,432$ in $\mathbf{W}^{h}$ to $131,574$ in $\mathbf{W}^h_l$ and $\mathbf{W}^h_r$. The computational complexity is significantly improved.

As a consequence, our proposed TD-ODE takes advantage of the continuous-time signal processing capability of ODE and the multi-domain structure of mmWave channels, so it is promising to achieve efficient continuous-time channel prediction, which will be demonstrated in the simulation section.

\subsection{Training and Testing Details}
	In this subsection, we supplement some training and testing details. To begin with, we adopt an offline training and online testing strategy. In the offline training stage, we use the clustered delay line (CDL) channel model to randomly generate $N_{\text{train}}$ time-varying channel samples. We divide these samples into $\frac{N_{\text{train}}}{BS}$ batches, with $BS$ being the batch size. We consider the $b$-th batch. 
	Each sample of this batch is a time-varying channel sequence, which is divided into two periods. The first period corresponds to the historical channels. In this period, we sample $J$ time slots with an equal time interval of $T_f$. The corresponding historical channels are $\mathbf{H}^{\text{input}} = \{\hat{\mathbf{H}}[-J + 1], \cdots, \hat{\mathbf{H}}[-1], \hat{\mathbf{H}}[0]\}$.
	The second period is regarded as the future time, where $P$ time slots are randomly sampled from the time duration $[0, KT_f]$. We use $t_1^b \le t_2^b \le \cdots \le t_P^b$ to index these sampled times in the $b$-th batch. Therefore, the corresponding noise-free channels are $\{\overline{\mathbf{H}}(t_1^b), \overline{\mathbf{H}}(t_2^b), \cdots, \overline{\mathbf{H}}(t_P^b) \}$, which are working as the training labels. Then, we use the proposed TN-ODE model to process $\mathbf{Y}^{\text{input}}$ and predict $\{\hat{\mathbf{H}}(t_1^b), \hat{\mathbf{H}}(t_2^b), \cdots, \hat{\mathbf{H}}(t_P^b) \}$. Finally, the NMSE is used for the loss function\footnote{\color{black} The reason NMSE is adopted as the loss function instead of MSE is that NMSE loss could speed up model convergence and avoid the influence of the amplitude of channel.}:
	\begin{align}
		\text{Loss} = \frac{1}{P}\sum_{i = 1}^{P} \mathbb{E}\left\{ \frac{\|\hat{\mathbf{H}}(t_i^b) - \overline{\mathbf{H}}(t_i^b) \|^2}{\|\overline{\mathbf{H}}(t_i^b) \|^2}\right\}.
	\end{align}
	Based on this loss function, the Adam optimizer is adopted to update the network parameters using their gradients. Notice that adjoint sensitivities proposed in \cite{ODE} are used to efficiently compute the ODE's gradients. The above procedure is carried out batch by batch until convergence. 

	The data size in the testing stage is $N_{\text{test}}$, where each channel sample is still divided into two periods. The first period is the same as that in the training stage. Regarding the second period, our target is to predict channels for future $KQ$ time slots (or $K$ frames). Therefore, we sample $KQ$ slots with an equal time interval of $T_s$, which are denoted by $t_i = i T_s$, $i = 1, \cdots, KQ$. Then, we use the well-trained TN-ODE model to predict $\hat{\mathbf{H}}(t_i)$, $i = 1, \cdots, KQ$. Finally, these predicted channels are used for precoding. 

{\color{black}
\subsection{Computational Complexity}
In this subsection, we provide a detailed computational complexity analysis of the proposed scheme in the testing stage. Here, we mainly count the number of complex-valued multiplications. 

For a sequence of historical channels $ \{\hat{\mathbf{H}}[-J + 1], \cdots, \hat{\mathbf{H}}[-1], \hat{\mathbf{H}}[0]\}$, the total $J$ channels are processed by the EncoderCell (\ref{eq:GRU2a})-(\ref{eq:GRU2d}) sequentially. Steps (\ref{eq:GRU2a})-(\ref{eq:GRU2c}) have a complexity in the order of $\mathcal{O}(F_lN_{\text{RF}}N_{\text{R}}M + F_l M F_r)$, and step (\ref{eq:GRU2d}) has a complexity of $\mathcal{O}(F_l F_r)$. Therefore, taking into account the $J$ channels, the computational complexity of the encoder is $\mathcal{O}(JF_lN_{\text{RF}}N_{\text{R}}M + JF_l M F_r) + \mathcal{O}(JF_l F_r) = \mathcal{O}(JF_lN_{\text{RF}}N_{\text{R}}M + JF_l M F_r)$.

As for the decoder, we can similarly derive that the computational complexities of calculating the functions DecoderCell$(\cdot)$ and Pred$(\cdot)$ are $\mathcal{O}(F_l^2 F_r + F_l F_r^2)$ and $\mathcal{O}( N_{\text{RF}} N_{\text{R}} F_l F_r + N_{\text{RF}} N_{\text{R}} F_r M)$, respectively. Moreover, the ODESolver($\cdot$) in (\ref{eq:ODESolver})  needs to calculate the DecoderCell$(\cdot)$ for $G$ times, where $G$ is proportional to $KQ$. Therefore, the computational complexity of the decoder is $\mathcal{O}(GF_l^2 F_r + GF_l F_r^2)$. Finally, as $KQ$ future channels are predicted, the overall number of complex-valued multiplications of the function Pred($\cdot$) is $\mathcal{O}( KQN_{\text{RF}} N_{\text{R}} F_l F_r + KQN_{\text{RF}} N_{\text{R}} F_r M)$.

As a consequence, the computional complexity of the proposed TN-ODE model is 
\begin{align}
\mathcal{O}( JF_lM (N_{\text{RF}}N_{\text{R}}+ F_r)) +
\mathcal{O}( GF_l F_r (F_l + F_r)) \notag\\
+
\mathcal{O}( KQN_{\text{RF}} N_{\text{R}}  F_r (F_l+ M)). 
\end{align}
}

\section{Simulation Results}\label{SIMULATION-RESULTS}
% \begin{table}[tbp]
% \centering
% \caption{\bf{Simulation Configurations}}
% \begin{tabular}{cccc}\hline
% \bf{Parameter} & \bf{Value}   & \bf{Parameter}       & \bf{Value} \\\hline
% $N_{\rm T}$     & 64      & $N_{\text{RF}}$ & 4\\
% $N_{\rm R}$     & 4       & $f_c$ & 28 GHz \\
% $B$ & 100 MHz & $M$  & 256\\
% SNR & 10 dB   & $N_{\text{train}}$ & 1000 \\
% $N_{\text{test}}$ & 200 & $T^{\text{prev}}$ & 0.625 ms \\
% $T^{\text{pred}}$ & 0.125 ms & $L^{\text{prev}}$ & 10 \\
% $L^{\text{pred}}_{\text{train}}$ & 5 & $L^{\text{pred}}_{\text{test}}$ & 10 \\
% $F_l$ & 64 & $F_r$ & 128 \\\hline
% \end{tabular}
% \label{tab1}
% \end{table}

\begin{table}[tbp]\color{black}
\centering
\caption{\color{black}\bf{Simulation Configurations}}
\begin{tabular}{cccc}\hline
\bf{Parameter} & \bf{Value}   & \bf{Parameter}       & \bf{Value} \\\hline
$N_{\rm T}$     & 128      & $N_{\text{RF}}$ & 4\\
$N_{\rm R}$     & 4       & $f$ & 28 GHz \\
$B$ & 100 MHz & $M$  & 256\\
SNR & 10 dB   & $N_{\text{train}}$ & 1000 \\
$N_{\text{test}}$ & 200 & $BS$ & 32 \\
$T_f$ & 0.625 ms & $T_s$ & 0.125 ms \\
$J$ & 10 & $P$ & 5 \\
$K$ &  2 & $Q$ & 5 \\
{$F_l$} & {64} & {$F_r$} & {128} \\\hline
\end{tabular}
\label{tab1}
\end{table}

In this section, simulation results are provided to demonstrate the superiority of our scheme. The CDL-B channel model in the Matlab 5G toolbox \cite{Transformer} is utilized to generate the data set. For each channel sample, the velocity of user is randomly generated from the uniform distribution $\mathcal{U}(30\:\text{km/h}, 60\:\text{km/h})$ and the delay spread is randomly chosen from the uniform distribution $\mathcal{U}(50\:\text{ns}, 200\:\text{ns})$. 
The simulation configurations are presented in Table \ref{tab1}. 
The compared benchmarks are as follows: 1) the perfect CSI; 2) the classical AI-based algorithms, including the GRU-based channel prediction~\cite{LSTM} and the FC network based algorithm~\cite{FC}; 3) the classical model-based techniques, including the prony-based angular-delay domain channel prediction (PAD)~\cite{04} and ST-AR~\cite{STAR} algorithms; 4) utilizing the outdated channels without prediction. 
% \begin{figure}[tp]
% 	\begin{center}
% 		\hspace*{0mm}\includegraphics[width=1\linewidth]{Rate.eps}
% 	\end{center}
% 	\vspace*{-4mm}
% 	\caption{Average rate performance against time slots.} 
% 	\label{fig:rate}
% 	\vspace*{+1mm}
% \end{figure} 

\begin{figure}
	\centering 
	%		\vspace*{-0.5em}
	\includegraphics[width=3.3in]{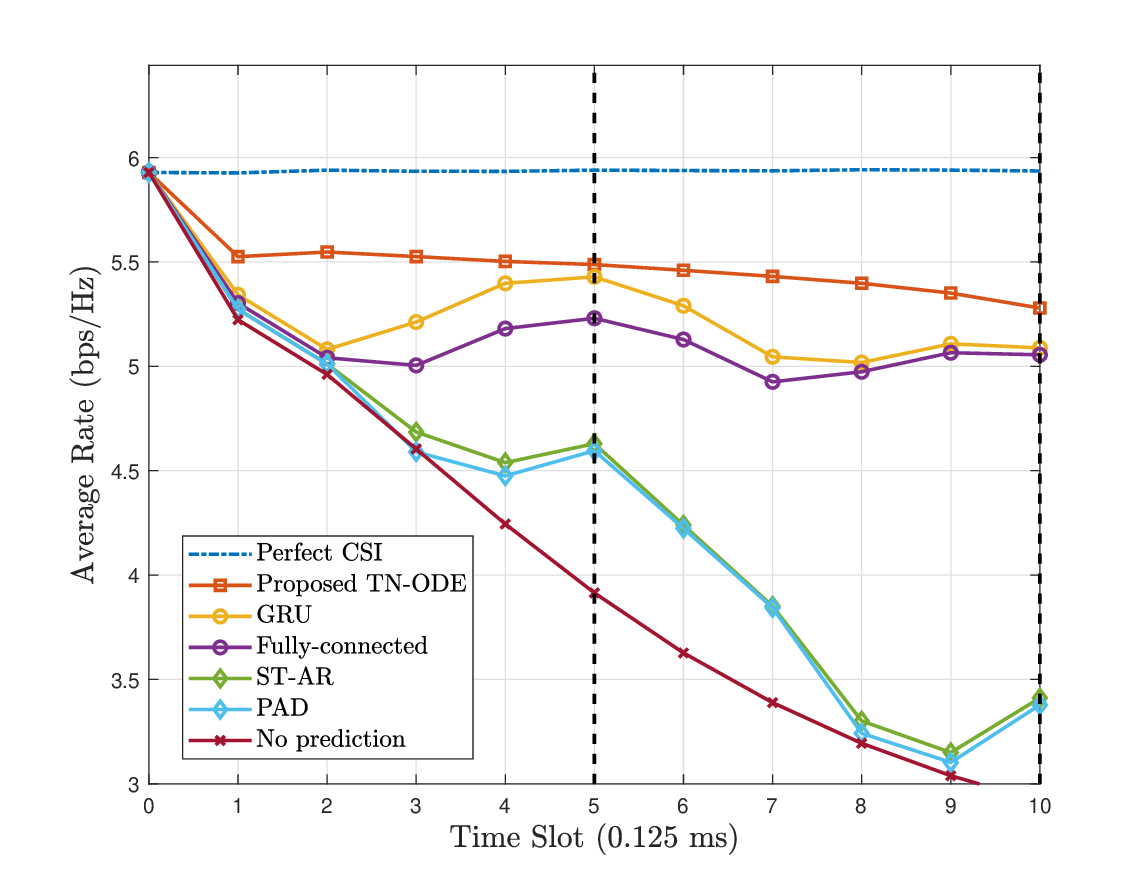}
	\caption{  Average rate performance against time slots.
	} 
	\label{fig:rate}
	\vspace*{-1em}
\end{figure}

In Figure \ref{fig:rate}, the average rate performance is evaluated. We follow the 5G standard and set $T_f$ as 0.625 ms and  $T_s$ as 0.125 ms. Therefore, the classical GRU, FC, ST-AR, and PAD algorithms predict the channels at the 5-th and 10-th time slots, and then recover the channels at other time slots through linear interpolation. 
It is clear from Figure \ref{fig:rate} that the average rate performance of classical algorithms degrades at the interpolated channels. Fortunately, our proposed scheme is able to avoid interpolation loss by predicting the future channels at all time slots with the assistance of TN-ODE. 
Additionally, the proposed TN-ODE exploits the multi-domain channel structure, so it can even achieve higher average rate than classical algorithms at the 5-th and 10-th time slots.

\begin{figure}\color{black}
	\centering 
	%		\vspace*{-0.5em}
	\includegraphics[width=3.3in]{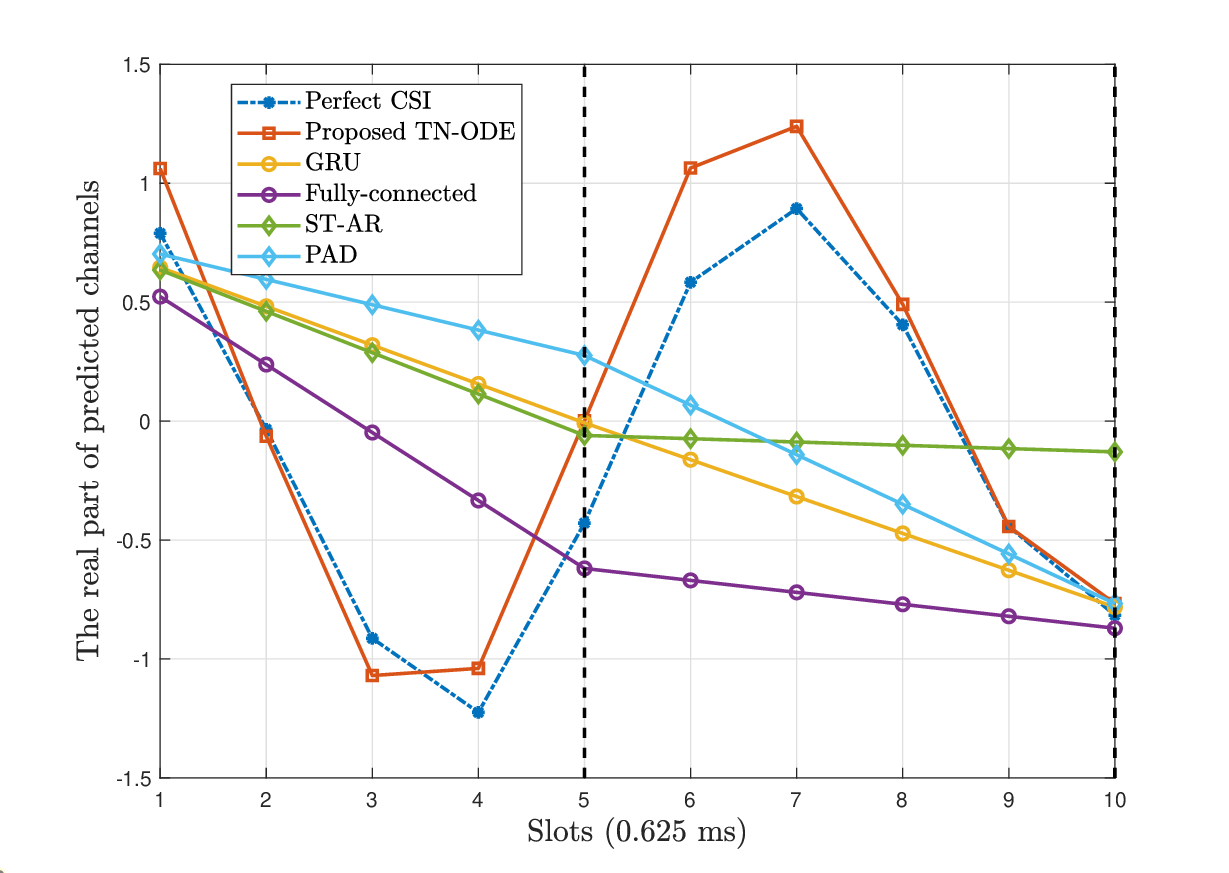}
	\caption{ \color{black} The real part of the true future channels and the predicted channels.
	} 
	\label{fig:shape}
	\vspace*{-1em}
\end{figure}

{\color{black} In Figure \ref{fig:shape}, the real part of the true future channels and the predicted channels for an arbitrary antenna index and } {\color{black}subcarrier are presented. We can observe from this figure that the existing discrete-time channel prediction techniques can only accurately predict the future channels at SRS positions, while the interpolated channels considerably deviate from the true channels. On the contrary, the proposed TN-ODE scheme well captures the dynamic of continuous-time channels. }
The simulation result in Figure \ref{fig:NMSE} further supports our discussion, where the NMSE performance against time slots is illustrated. It is obvious that the achieved NMSE of classical algorithms intensively fluctuates with respect to time slots, which is induced by the error of interpolation. On the contrary, the NMSE performance of our scheme  smoothly deteriorates over time, and it is always lower than -10 dB. As a result, we can conclude that our TN-ODE based approach accomplishes accurate continuous-time channel prediction.

% \begin{figure}[tp]
% 	\begin{center}
% 		\hspace*{0mm}\includegraphics[width=1\linewidth]{NMSE.eps}
% 	\end{center}
% 	\vspace*{-4mm}
% 	\caption{NMSE performance against time slots.} 
% 	\label{fig:NMSE}
% 	\vspace*{+1mm}
% \end{figure} 

\begin{figure}
	\centering 
	%		\vspace*{-0.5em}
	\includegraphics[width=3.3in]{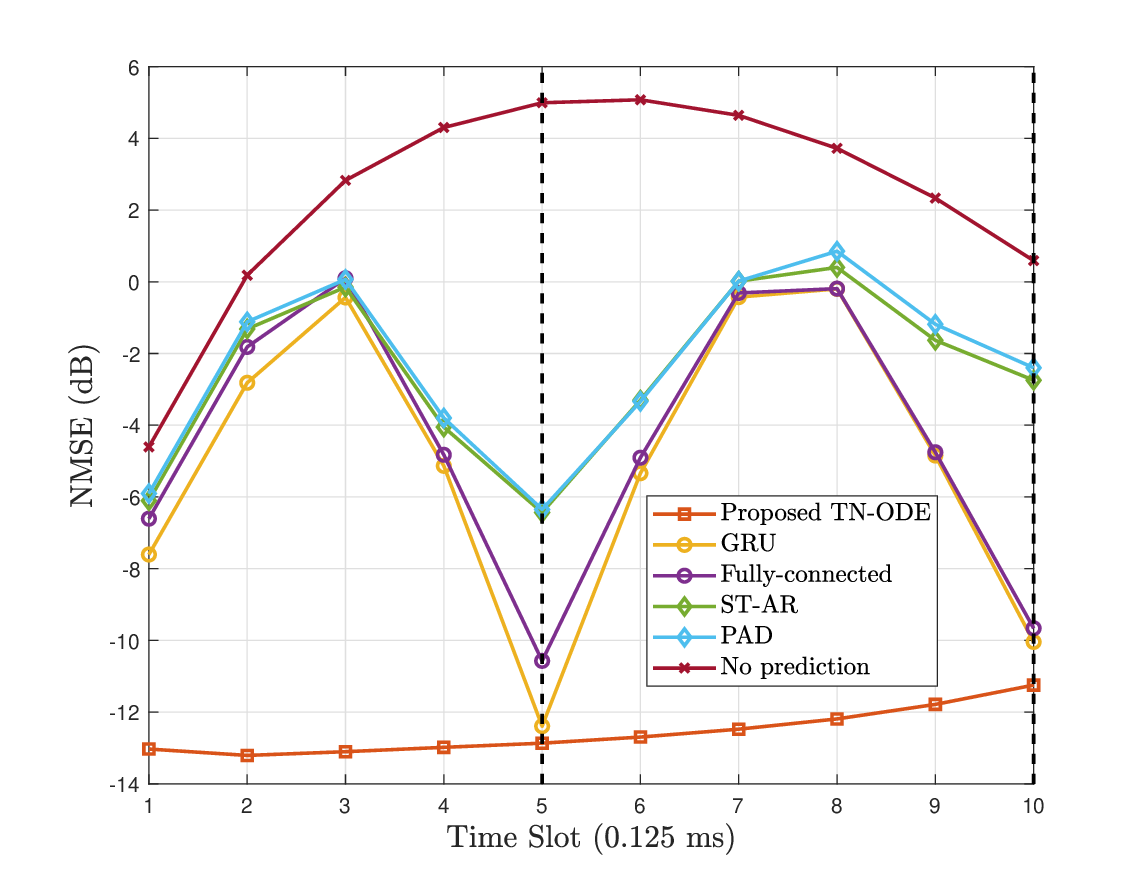}
	\caption{  NMSE performance against time slots.
	} 
	\label{fig:NMSE}
	\vspace*{-1em}
\end{figure}

}

\section{CONCLUSIONS}
\label{CONCLUSION}
In this paper, we have investigated the essential problem of continuous-time channel prediction in mobile mmWave massive MIMO systems. At first, we adopted the neural ODE to model the temporal correlation of mmWave channels, and then we introduced the neural ODE based channel prediction framework. This framework deployed a GRU-based encoder to extract features from historical channels and 
used a neural ODE based decoder to predict future continuous-time channels. After that, a TN-ODE model was proposed to improve this framework, which makes full use of the multi-domain channel structure. Simulations demonstrated that our scheme accomplished  accurate channel prediction in all time slots of several future frames.
The proposed TN-ODE model can be potentially extended to various continuous-time channel prediction scenarios, such as cell-free communication scenarios and RIS-aided communication scenarios. In the future, we will investigate the multi-user continuous-time channel prediction. 

	\balance
	\bibliographystyle{IEEEtran}
	\bibliography{IEEEabrv,myref}
	% that's all folks
\end{document}